\documentclass[]{article}
\usepackage[T1]{fontenc}
\usepackage{lmodern}
\usepackage{amssymb,amsmath}
\usepackage{ifxetex,ifluatex}
\usepackage{fixltx2e} % provides \textsubscript
\IfFileExists{upquote.sty}{\usepackage{upquote}}{}
\ifnum 0\ifxetex 1\fi\ifluatex 1\fi=0 % if pdftex
  \usepackage[utf8]{inputenc}
\else % if luatex or xelatex
  \ifxetex
    \usepackage{mathspec}
    \usepackage{xltxtra,xunicode}
  \else
    \usepackage{fontspec}
  \fi
  \defaultfontfeatures{Mapping=tex-text,Scale=MatchLowercase}
  
\fi
\IfFileExists{microtype.sty}{\usepackage{microtype}}{}
\usepackage[margin=1in]{geometry}
\usepackage{color}
\usepackage{fancyvrb}

\DefineVerbatimEnvironment{Highlighting}{Verbatim}{commandchars=\\\{\}}
\usepackage{framed}
\definecolor{shadecolor}{RGB}{248,248,248}
\newenvironment{Shaded}{\begin{snugshade}}{\end{snugshade}}
\newcommand{\KeywordTok}[1]{\textcolor[rgb]{0.13,0.29,0.53}{\textbf{{#1}}}}

\newcommand{\NormalTok}[1]{{#1}}
\usepackage{longtable,booktabs}
\usepackage{graphicx}
\makeatletter
\def\ScaleIfNeeded{%
  \ifdim\Gin@nat@width>\linewidth
    \linewidth
  \else
    \Gin@nat@width
  \fi
}
\makeatother
\let\Oldincludegraphics\includegraphics
{%
 \catcode`\@=11\relax%
 \gdef\includegraphics{\@ifnextchar[{\Oldincludegraphics}{\Oldincludegraphics[width=\ScaleIfNeeded]}}%
}%
\ifxetex
  \usepackage[setpagesize=false, % page size defined by xetex
              unicode=false, % unicode breaks when used with xetex
              xetex]{hyperref}
\else
  \usepackage[unicode=true]{hyperref}
\fi
\hypersetup{breaklinks=true,
            bookmarks=true,
            pdfauthor={JJ Merelo},
            pdftitle={Evolution of the number of GitHub users in Spain},
            colorlinks=true,
            citecolor=blue,
            urlcolor=blue,
            linkcolor=magenta,
            pdfborder={0 0 0}}
\usepackage{titling}
  \title{Evolution of the number of GitHub users in Spain}
  \pretitle{\vspace{\droptitle}\centering\huge}
  \posttitle{\par}
  \author{JJ Merelo}
  \preauthor{\centering\large\emph}
  \postauthor{\par}
  \predate{\centering\large\emph}
  \postdate{\par}
  \date{2/02/2016}

\begin{document}

\maketitle

\subsection{Abstract}\label{abstract}

\begin{quote}
Since we started measuring the community of GitHub users in Spain, it
has kept increasing in numbers in such a way that it has grown almost
50\%, to the current 12000, in less than one year. However, the reasons
for this are not clear. In this paper we will try to find out what are
the different components in this growth, or at least those that can be
measured, in order to find out which ones are due to the measurement
itself and which might be due to other reasons. In this paper we make an
advance towards finding out the reasons for this growth.
\end{quote}

\section{Introduction}\label{introduction}

Since the beginning of 2015, we have been measuring the number of users
of the popular open source software repository GitHub ({Merelo} et al.
2015,Merelo-Guervos et al. (2015)) in order to find out mainly what is
the geographical distribution of users and if there are real differences
across the country (Merelo 2015). Data was gathered by using GitHub
search API to retrieve users and then web scraping to retrieve user
data, such as the number of followers, contributions, and incorporation
date. This data is published periodically, ordered by province and
collated by autonomous region and the
\href{https://github.com/JJ/top-github-users-data/blob/master/formatted/top-alt-Spain.md}{whole
country}, in this case showing only the top 1000. These rankings are
shown in Twitter and other social media, with comments on who is going
up, down, and new shows.

\section{Methodology}\label{methodology}

The whole procedure has been described in ({Merelo} et al. 2015).
Periodically

\begin{itemize}
\itemsep1pt\parskip0pt\parsep0pt
\item
  We use GitHub API to search for \emph{active} users that have Spanish
  location in their profile.
\item
  Scraping is then used on the profile to extract data not returned by
  the API.
\item
  Data in JSON and CSV format is stored in a git repository, Markdown
  rankings are generated and published.
\end{itemize}

By ``periodically'' it means that we do it from time to time. It is
usually done every week, but it is also done in a domestic computer, so
some weeks is not really done and there are gaps. On the other hand,
collation of results to generate a single file with all users is done by
hand, so it is even less reliable. It is generally done every month,
though. All this data is published in the GitHub repository
\href{https://github.com/JJ/top-github-users-data}{JJ/top-github-users-data}.
Let us look at this data next.

\section{Numbers}\label{numbers}

The popularity of the rankings has been remarkable, and this has been
reflected in the number of users that these rankings show. The actual
numbers are shown in the following table

\begin{Shaded}
\begin{Highlighting}[]
\NormalTok{knitr::}\KeywordTok{kable}\NormalTok{(date_users)}
\end{Highlighting}
\end{Shaded}

\begin{longtable}[c]{@{}lr@{}}
\toprule
Date & Users\tabularnewline
\midrule
\endhead
2015-03-26 & 8659\tabularnewline
2015-04-09 & 8968\tabularnewline
2015-04-18 & 9496\tabularnewline
2015-04-19 & 9574\tabularnewline
2015-04-19 & 9574\tabularnewline
2015-04-22 & 9725\tabularnewline
2015-04-25 & 9808\tabularnewline
2015-05-01 & 9776\tabularnewline
2015-05-03 & 9870\tabularnewline
2015-05-07 & 9999\tabularnewline
2015-05-08 & 10024\tabularnewline
2015-05-09 & 10074\tabularnewline
2015-05-13 & 10157\tabularnewline
2015-05-27 & 10349\tabularnewline
2015-05-30 & 10418\tabularnewline
2015-05-31 & 10431\tabularnewline
2015-06-11 & 10539\tabularnewline
2015-06-13 & 10550\tabularnewline
2015-06-13 & 10557\tabularnewline
2015-06-14 & 10559\tabularnewline
2015-06-20 & 10625\tabularnewline
2015-06-23 & 10626\tabularnewline
2015-06-27 & 10674\tabularnewline
2015-07-05 & 10733\tabularnewline
2015-07-06 & 10722\tabularnewline
2015-07-22 & 10788\tabularnewline
2015-07-31 & 10854\tabularnewline
2015-08-17 & 10872\tabularnewline
2015-08-29 & 11037\tabularnewline
2015-09-06 & 11123\tabularnewline
2015-09-20 & 11253\tabularnewline
2015-10-03 & 11412\tabularnewline
2015-10-18 & 11660\tabularnewline
2015-10-24 & 11761\tabularnewline
2015-10-24 & 11756\tabularnewline
2015-10-25 & 11758\tabularnewline
2015-11-19 & 11975\tabularnewline
2015-11-26 & 11981\tabularnewline
2015-12-01 & 11993\tabularnewline
2015-12-11 & 12290\tabularnewline
2016-01-17 & 12530\tabularnewline
2016-01-23 & 12563\tabularnewline
2016-01-23 & 12527\tabularnewline
2016-01-23 & 12526\tabularnewline
2016-02-02 & 12849\tabularnewline
\bottomrule
\end{longtable}

From the initial 8659 users we have arrived lately at a number of users
in the vicinity of 13000. It is not believable, however, to claim that
this Big Bang of free software developers in Spain is singly due do this
ranking, or some co-occurring event. In fact, it might be due to several
different reasons.

\begin{itemize}
\itemsep1pt\parskip0pt\parsep0pt
\item
  Raw increase in the number of users: more people sign up for GitHub
  and start to contribute. This will happen naturally, independently of
  the published rankings.
\item
  Since the program uses the search API for identifying where the users
  are from, people who were already contributing fill out their profile
  so that they now show up in search results.
\item
  The rankings shows only users that have had non-zero contributions in
  the last year. Old users suddenly revisiting their old repositories
  will also start to show up in the rankings.
\end{itemize}

Our main concern with finding out the real reason is to check the actual
effect these rankings have had in the gamification of contributions to
open source, by creating a competition that will encourage more people
to sign in. However, it is not clear if this effect will actually
contribute to more users signing in or to current users being more
productive. That is why it is important to find out where new users
actually come from. So we will have to look at the distribution of users
per year of signing up.

\section{Distribution per year.}\label{distribution-per-year.}

The bar plot below shows the number of users by year of incorporation
into GitHub. This data is shown in the user profile.

\includegraphics{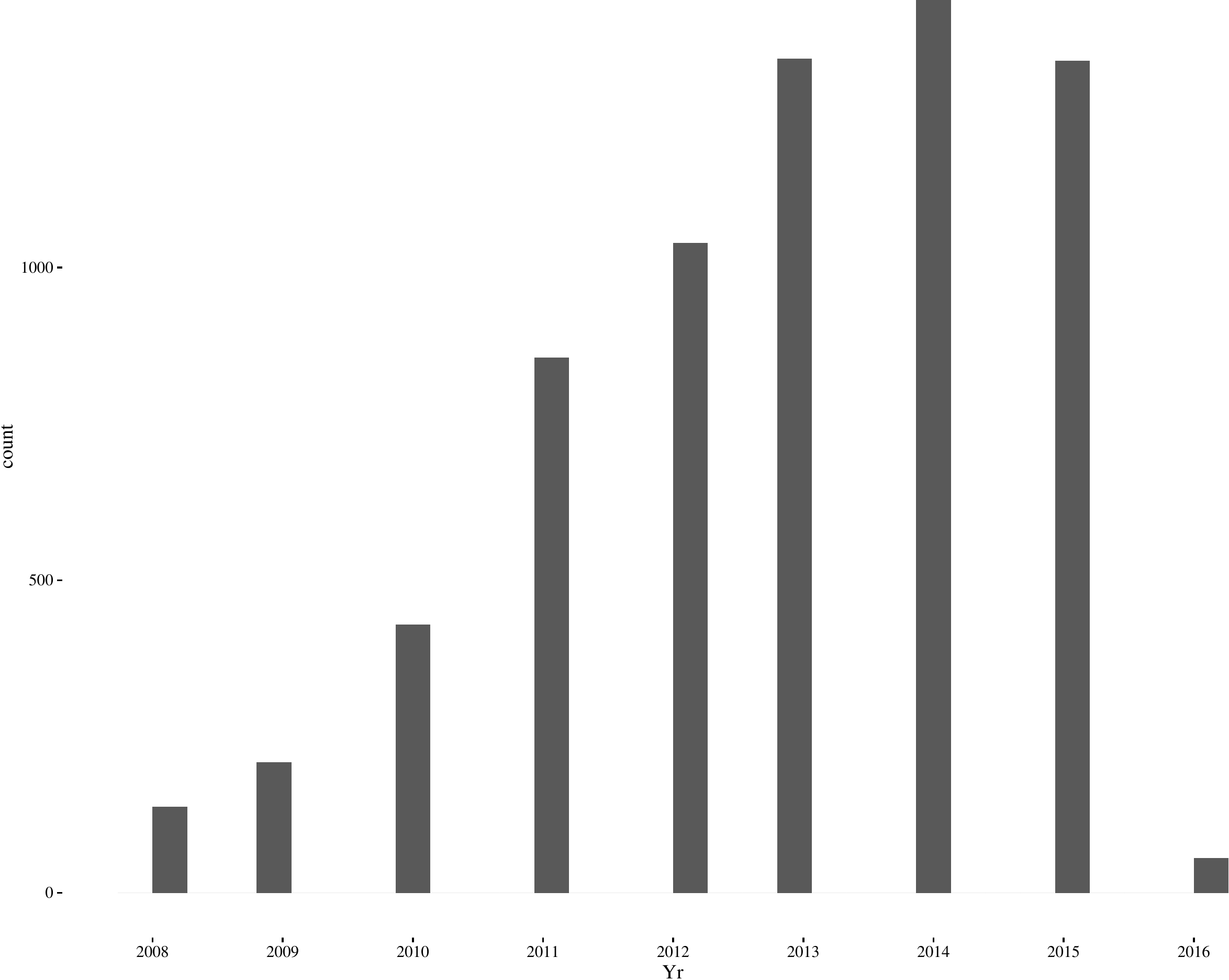}

Please bear in mind that this graph shows only the users that are
currently active, not those that might have been active in the year of
registration. And what it shows is that, in fact, users who signed up in
the previous year, are active in more quantities than those who signed
up during 2015. In fact, there were \emph{fewer} users registering last
year, the first year that we have been recording activity, than in 2014,
and approximately the same as in 2013.

In fact, users who have signed up last year account for only a small
fraction of total active users, less than 10\%. And, in fact, account
for less than 25\% of the new users that we have observed. This points
to the second hypothesis, users filling out their profile. But we will
delve further into this data in the next section

\section{\texorpdfstring{\emph{Really new} users and \emph{just new}
users.}{Really new users and just new users.}}\label{really-new-users-and-just-new-users.}

This means that we have got only around 1000 \emph{really new} users,
that is, users that have signed up during the period of measurement, and
these make up for a fraction of users just showing up. Let us break this
down per month in the next graph.

\includegraphics{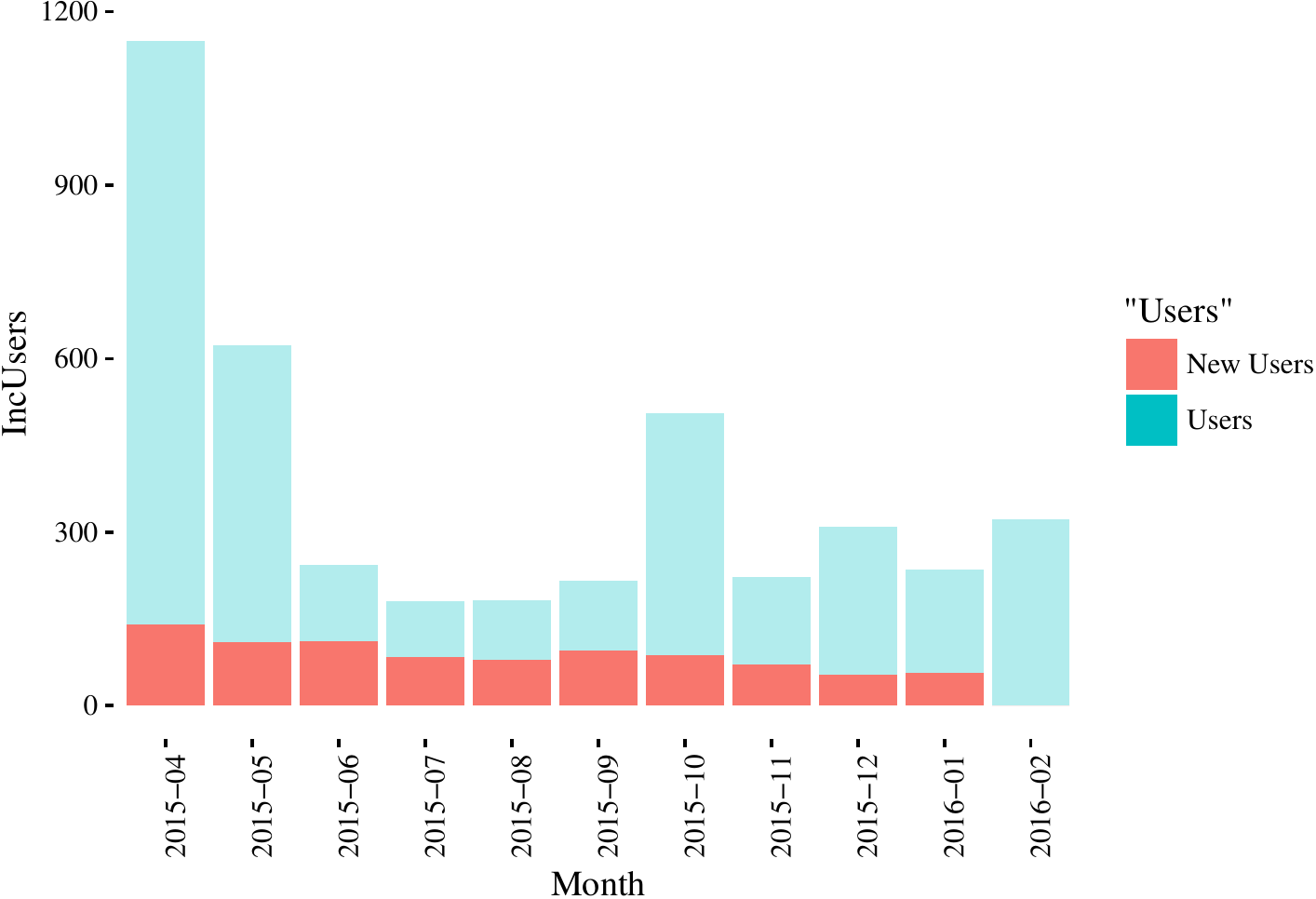} This
data has been extracted from the commit date, using git as a
time-stamped database. What we do is to extract all versions of the file
and use the \emph{last} every month to count the number of users
tracked. This number is then subtracted from the users we know about the
next month. This means that there might be a small time shift between
the number of new users (that are counted by the end of the month) and
the number of newly discovered users (which might have happened anytime
since the beginning of the month). This might be relevant if numbers
were closer, but since they are so different we do not think it makes a
difference. Anyway, people not counted one month are counted on the
next.

In this graph, \emph{new users} are those that have just signed up. As
it can be seen, they are a fraction of total users; while hew users
hover in the 100s and decrease towards the end of the year, total
accounted for users increase by around 300 every month, with the initial
months having a big boost due probably to the tuning of the search
strings and another boost in October probably due to the beginning of
the new university course and students and maybe new hires boasting
their presence in the rankings and encouraging their colleagues to
update their profile and show up too.

At any rate, it is quite clear that the ranks of new users increase
mainly \emph{not} due to \emph{really} new users, that is, persons newly
signing up and using GitHub.

\section{Conclusions}\label{conclusions}

In this work, one of our objectives was to study the impact of measuring
the activity of GitHub users in Spain and do it periodically, observing
changes, patterns and also geographical trends. We expected that this
would contribute to open source in general, since GitHub is nowadays the
repo container of choice for FLOSS developers. It is quite clear that
there is an impact, at least if we consider the number of contributions.
However, the impact is mainly \emph{not} in the number of users newly
signing up for GitHub, but on users either completing their profile or
making contributions after staying dormant for some time. We do not know
which one will be, but an educated guess is that probably, over all due
to increments during academically important periods, might be the
latter: students coming back to their repos when the school year starts;
this could account for the big bump shown in October 2015. However, we
will have to continue measurements during several years to actually make
some affirmation. The fact that the total number of users does not
decrease, since it measures activity \emph{only} in the last year, is
maybe supporting this hypothesis, which, on the other hand, is almost
impossible to check since the date of change of profile data is not
known.

On the other hand, we are more interested in measuring other ways this
ranking is changing the community as future lines of works. Examining
particular provinces like Granada, for instance, or examining
collaboration graphs. That is left as future work.

\section{Notes and acknowledgements}\label{notes-and-acknowledgements}

This report has been supported by GeNeura team and its supporting grants
such as the project TIN2014-56494-C4-3-P (Spanish Ministry of Economy
and Competitiveness), funded by the Spanish Ministry of Economy and
Competitivity and FEDER. \href{http://geneura.wordpress.com}{GeNeura}
supports open science, and this paper is written in RMarkdown and its
source, supporting data and processing scripts are available from GitHub
at \url{https://github.com/JJ/gh-in-spain-2016}. Links to data and
presentation are included in this web page
\url{http://jj.github.io/gh-in-spain-2016/}.

\section*{References}\label{references}
\addcontentsline{toc}{section}{References}

{Merelo}, J. J., N. {Rico}, I. {Blancas}, M. G. {Arenas}, F. {Tricas},
and J. A. {Vacas}. 2015. ``Measuring the Local GitHub Developer
Community.'' \emph{ArXiv E-Prints}, January.

Merelo, Juan J. 2015. ``GitHub Users in Spain: An Geospatial Analysis,''
April. \url{http://dx.doi.org/10.6084/m9.figshare.1384884}.

Merelo-Guervos, Juan Julián, Israel Blancas, M. G. Arenas, Fernando
Tricas, José Antonio Vacas, and Nuria Rico. 2015. ``GitHub Rankings and
Its Impact on the Local Free Software Development Community.'' \emph{The
Winnower}, January.
doi:\href{http://dx.doi.org/10.15200/winn.142251.14740}{10.15200/winn.142251.14740}.

\end{document}